\documentclass[useAMS,usenatbib,onecolumn]{mn2e}
\usepackage{graphics}
\usepackage{times}
\def\be{\begin{equation}}
\def\ee{\end{equation}}

\begin{document}

\title[Likelihood techniques for CMB power spectra]{Likelihood
techniques for the combined analysis of CMB 
temperature and polarisation power spectra.} 
\author[W.J. Percival \& M.L. Brown]
{W.J. Percival$^{1}$\thanks{E-mail: will.percival@port.ac.uk, 
    mlb@roe.ac.uk} \& M.L. Brown$^{2\,\star}$\\
$^{1}$ Institute of Cosmology and Gravitation, University of
       Portsmouth, Portsmouth, P01 2EG\\
$^{2}$ Institute for Astronomy,
       University of Edinburgh, Blackford Hill, Edinburgh EH9 3HJ}

\oddsidemargin 0.0in                    
\textheight 25.0cm                      
\textwidth 17cm                         
\topmargin -2.0cm                        

\date{\today} 

\maketitle

\begin{abstract}

  We consider the shape of the likelihood and posterior surfaces to be
  used when fitting cosmological models to CMB temperature and
  polarisation power spectra measured from experiments. In the limit
  of an all-sky survey with Gaussian distributed pixel noise we show
  that the true combined likelihood of the four CMB power spectra (TT,
  TE, EE \& BB) has a Wishart distribution and we discuss the
  properties of this function.  We compare various fits to the
  posterior surface of the $C_l$s, both in the case of a single
  auto-power spectrum and for a combination of temperature and
  polarisation data. In the latter case, it is important that the fits
  can accurately match the Wishart distribution in the limit of near
  full-sky coverage. Simple extensions of auto-power spectrum fits to
  include polarisation data generally fail to match correlations
  between the different power spectra in this limit. Directly fitting
  pixel values on large scales, as undertaken by the WMAP team in
  their analysis of the 3 year data, avoids the complications of
  characterising the shape of the posterior for the power
  spectra. Finally we demonstrate the importance of the likelihood
  distribution on analytic marginalisation, and provide a formula for
  analytic marginalisation over a calibration error given an all-sky
  survey.

\end{abstract}


\section{Introduction}

In this era of precision cosmology \citep{spergel03,spergel06}, it is
vital that care is taken with every step involved in the analysis and
interpretation of cosmological data. In this paper we consider the
likelihood technique used to fit cosmological models to CMB power
spectra. For their analysis of the 3-year data
\citep{hinshaw06,page06}, the WMAP team have adopted a pixel-based
likelihood analysis at low multipoles, thus avoiding the complications
introduced by fitting to the shape of the posterior surface
\citep{slosar04}. This posterior surface is strongly non-Gaussian
which must be accounted for when performing model comparisons using
power spectra.  In addition, there are now important constraints on
both the temperature and polarisation power spectra, which are not
independent and need to be jointly analysed.

In this paper we review previous work analysing the posterior surface
for temperature power spectra and extend this to include polarisation
data. Initially, we present exact formulae for all-sky surveys with
negligible noise. The inclusion of isotropic uncorrelated Gaussian
distributed pixel noise will not change the form of the posterior
surface for the $C_l$s as it will simply increase the variance in the
$a_{lm}$s -- to calculate the $C_l$s we are still summing the squares
of Gaussian random variables. However, an incomplete sky map will
change the posterior distribution, causing correlations between modes,
and will also change the posterior shape. For any survey, the central
limit theorem can be invoked to show that at large $l$, the likelihood
distribution will tend to a multi-variate Gaussian form. We therefore
see that the true likelihood will interpolate between a skewed
distribution on large scales, and a multi-variate Gaussian
distribution on small scales. Fitting formulae which are able to match
this intrinsic change in shape have previously been adopted to provide
an approximate likelihood calculation for a single auto-power
spectrum. The primary aim of our paper is to extend these fits to the
combination of temperature and polarisation data.

The layout of our paper is as follows. In Section~\ref{sec:like} we
lay the groundwork by briefly reviewing the standard Bayesian approach
to model selection, using Bayes theorem to link the likelihood to the
posterior of interest. In Section~\ref{sec:all-sky-temp} we review the
well known likelihood and posterior distributions for the TT power
spectrum of an all-sky noiseless survey. This work is expanded in
Section~\ref{sec:wishart} to include polarisation data in the same
limit of no noise and an all-sky survey -- the combined likelihood is
given by the Wishart distribution.  The important properties,
including the multi-variate Gaussian limit at large $l$ are
presented. In Section~\ref{sec:complications} we discuss the
complications introduced by realistic surveys on the posterior
shape. These complications have led previous studies to consider
posterior fits for the TT power spectrum, the most common of which is
the log-normal distribution \citep{bond00}. A number of these fits to
the posterior surface are compared in
Section~\ref{sec:posterior_fit_single}. The extension of these fits to
the joint analysis of temperature and polarisation data is considered
in Section~\ref{sec:posterior_fit_multi}. The importance of this work
setting the mathematical foundation for the shape of the posterior
surface is demonstrated in Section~\ref{sec:marginalise}, where we
consider marginalising over a Gaussian distributed calibration error
in the TT power spectrum. The difference between assuming a Gaussian
posterior, commonly used in the literature, and using the correct
distribution for an all-sky survey is demonstrated. We discuss our
results in Section \ref{sec:discuss}.

\section{likelihoods and posteriors} \label{sec:like}

Although it is probably the most reprinted equation in scientific
literature, it is central to the work presented here, so for
completeness we include the standard Bayesian equation
\be
  f({\bf X}|{\bf \hat{X}}) = \frac{f({\bf \hat{X}}|{\bf X})f({\bf X})}
                                  {f({\bf \hat{X}})},
  \label{eq:bayes}
\ee
which relates the likelihood function $f({\bf \hat{X}}|{\bf X})$ --
the distribution of the data ${\bf \hat{X}}$ given a model ${\bf X}$,
to the posterior $f({\bf X}|{\bf \hat{X}})$ -- the distribution of
models given the data. The prior $f({\bf X})$, which cannot be
avoided, provides the information that we already know about the
models. For example, our prior might be that only 6-parameters are
needed to model present CMB data, and that we initially know nothing
about those parameters -- they themselves have uniform priors.  It is
probably worth emphasising that the distributions of the likelihood
and posterior can have different forms. For example, if the likelihood
is Gaussian, but has a variance that depends on the model, then the
posterior will not be Gaussian, for a uniform prior on the models. We
must therefore take care to distinguish likelihood and posterior.

In this work, we will quantify the data ${\bf \hat{X}}$ using the
power spectra measured from some experiment, and the models ${\bf X}$
by the same statistic.  For an all-sky survey, where different modes
are independent, the likelihood is 
\be 
  f({\bf \hat{X}}|{\bf X}) = \prod_l f(\hat{C}_l^{\rm XX}|C_l^{\rm XX}), 
\ee 
where $\hat{C}_l^{\rm XX}$ represents the measured power spectra,
$C_l^{\rm XX}$ the model power spectra, and ${\rm XX} = {\rm
  TT,EE,TE,BB}$. The posterior that we are interested in, $f({\bf
  X}|{\bf \hat{X}})$ is dependent on the product of the likelihoods of
individual multipoles. For a single multipole, we see that it is the
dependence of the likelihood on the model power spectrum ($C_l^{\rm
  XX}$) that is of interest. In this paper we refer to $f(C_l^{\rm
  XX}|\hat{C}_l^{\rm XX})$ as the posterior for the power spectra, as
it is related to $f(\hat{C}_l^{\rm XX}|C_l^{\rm XX})$ using Bayes
theorem and a uniform prior on $f(C_l^{\rm XX})$. For a given
experiment, it is the posterior $f(C_l^{\rm XX}|\hat{C}_l^{\rm XX})$
that tells us the model constraints provided by the data on a
particular scale.

It is worth emphasising that a cosmological likelihood analysis based
on the work presented in our paper does not depend on the assumption
of a uniform prior in the power spectra. Instead, in such an analysis,
the prior would be defined by the set of allowed models. In this paper
we do not consider a specific set of cosmological models, and simply
focus on the dependence of the likelihood on our chosen model
statistic -- the model power spectrum $C_l^{\rm XX}$. We have decided
to call this the posterior, simply because it can be considered in
this way following the assumption of a uniform prior in $f(C_l^{\rm
  XX})$. Had we quantified the models by a different statistic,
$h(C_l^{\rm XX})$, then the prior on the allowed set of cosmological
models $f({\bf X})$ is unchanged -- if we considered a grid of
cosmological models, then this grid is unchanged. The likelihood for
each $l$ would have to be multiplied by $dh(\hat{C}_l^{\rm
  XX})/d\hat{C}_l^{\rm XX}$ to allow for the change of variables, but
we see that this is independent of the model values. Consequently, as
we would hope, the same posterior distribution for the set of
cosmological models would be recovered whatever statistic is used to
quantify the models, provided there is no loss of information
associated with this choice.

\section{Likelihood and posterior distributions for all-sky noiseless surveys}
\label{sec:likepost-all-sky}
\subsection{Temperature only data}  \label{sec:all-sky-temp}

In an all-sky, noiseless CMB survey, the Spherical Harmonic
coefficients $a_{lm}^{\rm T}$ of the temperature fluctuations obey a 
Gaussian distribution
\be
  f(a_{lm}^{\rm T}|C_l^{\rm TT}) = \frac{1}{\sqrt{2\pi C_l^{\rm TT}}} 
     \exp\left[ \frac{-|a_{lm}^{\rm T}|^2}{2C_l^{\rm TT}}\right],
  \label{eq:stdgauss}
\ee
where $C_l^{\rm TT} = \langle |a_{lm}^{\rm T}|^2 \rangle$. 
Using statistical isotropy, we can average over $m$ and define an
estimator of the power
\be
  \hat{C}_l^{\rm TT} = \frac{1}{2l+1}\sum_m|a_{lm}^{\rm T}|^2.
  \label{eq:hatcl}
\ee

The sum of the squares of $\nu$ standard Gaussian random variables has
a $\chi^2$ distribution with $\nu$ degrees of freedom. Consequently,
$\hat{Y}_l=\sum_m|a_{lm}^{\rm T}/\sqrt{C_l^{\rm TT}}|^2$ will have a
$\chi^2$ distribution with $\nu=2l+1$ degrees of freedom
\be
  f(\hat{Y}_l|C_l^{\rm TT}) = 
    \frac{\hat{Y}_l^{\nu/2-1}}{\Gamma(\nu/2)2^{\nu/2}}
    \exp\left[-\frac{\hat{Y}_l}{2}\right].
\ee

$\hat{C}_l^{\rm TT}$ is a multiple of $\hat{Y}_l$, $\hat{C}_l^{\rm
TT} = C_l^{\rm TT} \hat{Y}_l /\nu$ so the likelihood will
have a $\Gamma$ distribution
\be
  f(\hat{C}_l^{\rm TT}|C_l^{\rm TT}) 
    \propto C_l^{\rm TT}\left( \frac{\hat{C}_l^{\rm TT}} 
      {C_l^{\rm TT}} \right)^{\nu/2-1}
    \exp\left[-\frac{\nu\hat{C}_l^{\rm TT}}{2C_l^{\rm TT}}\right].
  \label{eq:fCl_gamma}
\ee 
The mean and variance of this distribution are $C_l^{\rm TT}$ and
$2(C_l^{\rm TT})^2/\nu$. Note that the maximum of this distribution
occurs at $\hat{C}_l^{\rm TT}=(\nu-2)/\nu C_l^{\rm TT}$, not at the
mean value. In the limit $\nu\to\infty$, the central limit theorem
gives that this distribution tends towards a Gaussian with the same
mean and variance.

Using Bayes theorem to convert to the posterior $f(C_l^{\rm
TT}|\hat{C}_l^{\rm TT})$, assuming a uniform prior in
$f(C_l^{\rm TT})$ gives
\be
  f(C_l^{\rm TT}|\hat{C}_l^{\rm TT}) 
    \propto (C_l^{\rm TT})^{-\nu/2}
    \exp\left[-\frac{\nu\hat{C}_l^{\rm TT}}{2C_l^{\rm TT}}\right],
  \label{eq:Cl_posterior}
\ee
where we have not included the normalisation factor dependent on
$\hat{C}_l^{\rm TT}$. The maximum of this function
occurs at $\hat{C}_l^{\rm TT}$, while the mean is given by 
\be
  \langle C_l^{\rm TT}\rangle=\frac{\nu}{\nu-8}\hat{C}_l^{\rm TT}.
\ee
This offset between maximum and mean is simply a result of the
skewness of the distribution. As $\nu\to\infty$, this distribution will
tend towards a Gaussian form.  This follows from the Bayesian identity
and the fact that $f(\hat{C}_l^{\rm TT}|C_l^{\rm TT})$ tends to a
Gaussian distribution (see the discussion following equation
(\ref{eq:L_epsilon_true})).

We can take the logarithm of the posterior to give
\be
  -\ln f(C_l^{\rm TT}|\hat{C}_l^{\rm TT}) = \frac{\nu}{2}
    \left(\ln C_l^{\rm TT} 
          + \hat{C}_l^{\rm TT} / C_l^{\rm TT} \right),
  \label{eq:Lallsky}
\ee 
ignoring an irrelevant additive constant. From this, we can calculate
the curvature around the distribution maximum
\be
  -\left.\frac{d^2 \ln f(C_l^{\rm TT}|\hat{C}_l^{\rm TT})}{d^2C_l^{\rm TT}}
    \right|_{\hat{C}_l^{\rm TT}} \propto
    \frac{1}{(\hat{C}_l^{\rm TT})^2}.
  \label{eq:fisher1}
\ee

There is an alternative way of deriving equation (\ref{eq:Lallsky}) by
considering the joint probability density of $a_{lm}^{\rm T}$ for
$-l\le m\le l$, 
\be
  f(a_{l,m=-l}^{\rm T},...,a_{l,m=l}^{\rm T} |C_l^{\rm TT}) = 
    \prod_m f(a_{lm}^{\rm T}|C_l^{\rm TT}).
\ee
Substituting for $f(a_{lm}^{\rm T}|C_l^{\rm TT})$ from equation
(\ref{eq:stdgauss}), this reduces to 

\be
  - \ln f(C_l^{\rm TT}|\hat{C}_l^{\rm TT}) = \frac{1}{2} 
    \sum_m \left(\ln C_l^{\rm TT}
                 + |a_{lm}^{\rm T}|^2 / C_l^{\rm TT}\right),
\ee
where once again, we have ignored an irrelevant additive component. 
We see that this distribution is only dependent on $a_{lm}^{\rm T}$ 
through $\hat{C}_l^{\rm TT}$, and that substituting in 
$\hat{C}_l^{\rm TT}$ from equation (\ref{eq:hatcl}) leads to the 
same $C_l^{\rm TT}$ dependence that we had in equation
(\ref{eq:Lallsky}). 

\subsection{Including polarisation data}
  \label{sec:wishart}

If we now include E-mode and B-mode polarisation data, there are 3
spherical harmonic coefficients of interest, $a_{lm}^{\rm T}$,
$a_{lm}^{\rm E}$ and $a_{lm}^{\rm B}$. These are multivariate Gaussian
random variables with expected values of zero. The data vector ${\bf
X}_a$ and covariance matrix ${\bf V}_l$ for the multivariate Gaussian
can be written as
\be
  \bf{X}_a = \left(\begin{array}{c} 
    a_{lm}^{\rm T} \\
    a_{lm}^{\rm E} \\
    a_{lm}^{\rm B} \end{array}
  \right),
\,\,\,\,\,
  {\bf V}_l = \left(\begin{array}{ccc} 
    C_l^{\rm TT} & C_l^{\rm TE} & 0 \\
    C_l^{\rm TE} & C_l^{\rm EE} & 0 \\
    0            & 0            & C_l^{\rm BB} \end{array}
  \right).
\ee
Note that the cross-correlations between different parity fields are
expected to be zero (and $B$ has the opposite parity to $E$ and $T$).
The random variables of interest are the elements of the matrix
\be
  {\bf S}_l = \frac{1}{2l+1}\sum_m {\bf X}_a {\bf X}^\dagger_a
    = \left(\begin{array}{ccc} 
    \hat{C}_l^{\rm TT} & \hat{C}_l^{\rm TE} & \hat{C}_l^{\rm TB} \\
    \hat{C}_l^{\rm TE} & \hat{C}_l^{\rm EE} & \hat{C}_l^{\rm EB} \\
    \hat{C}_l^{\rm TB} & \hat{C}_l^{\rm EB} & \hat{C}_l^{\rm BB} 
    \end{array}\right),
  \label{eq:Sl}
\ee
where ${\bf X}^\dagger_a$ represents the Hermitian conjugate of ${\bf
  X}_a$. The matrix ${\bf S}_l$ has a Wishart distribution with
$\nu=(2l+1)$ degrees of freedom. There is a slight complication caused
by defining ${\bf S}_l$ as the average over the $(2l+1)$ modes rather
than the sum, which is used to derive the standard Wishart
distribution. ${\bf S}_l$ still has a Wishart distribution, but we
need to consider the matrix ${\bf W}_l={\bf V}_l/(2l+1)$ rather than
${\bf V}_l$.  The Wishart distribution is given by
\be
  f({\bf S}_l|{\bf W}_l) = 
    \frac{|{\bf S}_l|^{(\nu-p-1)/2}
    \exp\left[-{\rm trace}({\bf W}_l^{-1}{\bf S}_l/2)\right]}
    {2^{p\nu/2}|{\bf W}_l|^{\nu/2}\Gamma_p(\nu/2)},
  \label{eq:wishart}
\ee 
where ${\bf S}_l$ and ${\bf W}_l$ are positive definite
symmetric $p\times p$ matrices, $\nu>p$, and $\Gamma_p(\nu/2)$ is the
multi-variate Gamma function,
\be
  \Gamma_p(\nu/2) = \pi^{p(p-1)/4}
    \prod_{i=1}^{p}\Gamma\left[(\nu+1-i)/2\right].
\ee

Because of the form of ${\bf V}_l$, we can decompose into two Wishart
distributions, one with $p=2$ for $\hat{C}_l^{\rm TT}$,
$\hat{C}_l^{\rm EE}$, $\hat{C}_l^{\rm TE}$ and a separate,
independent, $p=1$ Wishart distribution for $\hat{C}_l^{\rm BB}$,
which reduces to a $\Gamma$ distribution as described in the previous
Section. To help to understand the form of the Wishart distribution,
we now explain how it is normalised, focusing on the $p=2$
distribution covering $\hat{C}_l^{\rm TT}$, $\hat{C}_l^{\rm EE}$ and
$\hat{C}_l^{\rm TE}$. The probability density function for ${\bf S}_l$
is equivalent to the joint distribution of the elements of the matrix,
over all positive definite matrices so
\be
  \int_0^\infty d\hat{C}_l^{\rm TT}
  \int_0^\infty d\hat{C}_l^{\rm EE}
  \int_{\sqrt[-]{\hat{C}_l^{\rm TT}\hat{C}_l^{\rm EE}}}
      ^{\sqrt[+]{\hat{C}_l^{\rm TT}\hat{C}_l^{\rm EE}}}
      d\hat{C}_l^{\rm TE}
  f({\bf S}_l|{\bf V}_l) = 1.
\ee

As expected, the marginal distributions of the diagonal elements of
${\bf S}_l$ have a $\Gamma$ distribution as described in
Section~\ref{sec:all-sky-temp}. However, the same is not true for the
off-diagonal elements. First, suppose that we have obtained
$\hat{C}_l^{\rm TT}$ and $\hat{C}_l^{\rm TE}$, but for some reason not
$\hat{C}_l^{\rm EE}$ (as for example with the WMAP year 1 data
\citealt{bennett03,hinshaw03}). The joint likelihood of
$\hat{C}_l^{\rm TT}$ and $\hat{C}_l^{\rm TE}$ can be obtained by
integrating equation (\ref{eq:wishart}), with $p=2$, over $\hat{C}_l^{\rm
EE}$ forcing ${\bf S}_l$ to be positive definite. The marginal
distribution of $\hat{C}_l^{\rm TT}$ with the constraint
$0<\hat{C}_l^{\rm TT}<\infty$, and $\hat{C}_l^{\rm TE}$ with the
constraint $-\infty<\hat{C}_l^{\rm TE}<\infty$ is
\be
  f(\hat{C}_l^{\rm TT},\hat{C}_l^{\rm TE}|
    C_l^{\rm TT},C_l^{\rm TE},C_l^{\rm EE}) = 
   \frac{\nu^{(\nu+1)/2}}{\sqrt{\pi}\Gamma(\nu/2)2^{\frac{\nu-1}{2}}|V|^{1/2}}
    \frac{(\hat{C}_l^{\rm TT})^{(\nu-3)/2}}
               {(C_l^{\rm TT})^{(\nu-1)/2}}
    \exp\left[-\frac{\nu}{2|V|}\left(\hat{C}_l^{\rm TT}C_l^{\rm EE}
              -2\hat{C}_l^{\rm TE}C_l^{\rm TE}
	      +\frac{(\hat{C}_l^{\rm TE})^2C_l^{\rm TT}}
                    {\hat{C}_l^{\rm TT}}\right)\right].
  \label{eq:tt-te_marg_like}
\ee
It is interesting to note that a constraint on $\hat{C}_l^{\rm TT}$
and $\hat{C}_l^{\rm TE}$ still leaves us with a likelihood that is
dependent on $C_l^{\rm EE}$ -- information is retained in
this case. In fact, were we presented with an all-sky survey with
negligible noise, then the likelihood analysis of any two or more of
the three possible $E$ and $T$ mode auto- and cross-power spectra
should be attempted using the full matrix ${\bf V}$

By integrating equation (\ref{eq:tt-te_marg_like}) over $\hat{C}_l^{\rm
TT}$, we obtain the following marginal distribution for
$\hat{C}_l^{\rm TE}$
\be
  f(\hat{C}_l^{\rm TE}|
    C_l^{\rm TT},C_l^{\rm TE},C_l^{\rm EE}) = 
    \frac{\nu}{\sqrt{\pi}\Gamma(\nu/2)2^{\frac{\nu-1}{2}}|V|^{1/2}}
    \left[\frac{(\nu\hat{C}_l^{\rm TE})^2}
      {C_l^{\rm TT}C_l^{\rm EE}}\right]^{(\nu-1)/4}
    \exp\left(\frac{\nu\hat{C}_l^{\rm TE}C_l^{\rm TE}}{|V|}\right)
    {\rm K}_{(\nu-1)/2}\left(
      \frac{\nu|\hat{C}_l^{\rm TE}|\sqrt{C_l^{\rm TT}C_l^{\rm EE}}}
         {|V|}\right),
  \label{eq:C_TE_marg}
\ee
where ${\rm K}_n$ is a modified Bessel function of the second kind,
and $|V|=C_l^{\rm TT}C_l^{\rm EE}-(C_l^{\rm TE})^2$. This marginal
distribution covers the interval $-\infty<\hat{C}_l^{\rm
TE}<\infty$. Again it is worth emphasising that the likelihood is
dependent on $C_l^{\rm TT}$ and $C_l^{\rm EE}$ in addition to
$C_l^{\rm TE}$. In a Bayesian analysis, the model constraint
provided by a measurement of $\hat{C}_l^{\rm TE}$ depends on all of
these model values.

\begin{figure*}
  \centering
  \resizebox{0.90\textwidth}{!}{  
    \rotatebox{-90}{\includegraphics{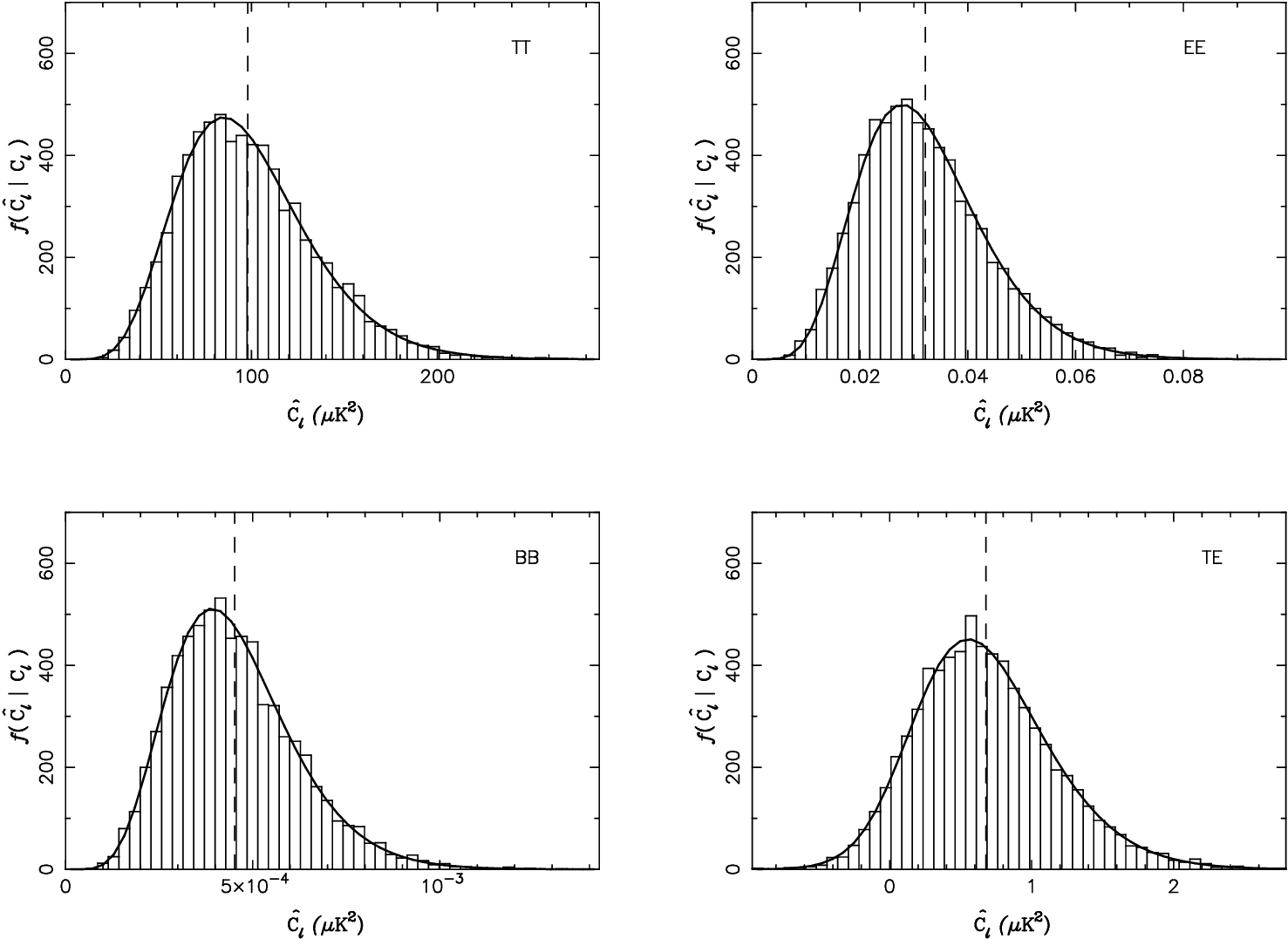}}}
  \caption{The distribution of recovered $\hat{C}_{l=7}$ calculated
  for the best-fit cosmological model to the 1-year WMAP data
  (histograms). A B-mode polarisation component was included in the simulations
  using a tensor-to-scalar ratio of $T/S = 0.05$. The recovered means,
  which coincide with the input model spectra, are shown as the
  vertical dashed lines. These distributions are shown to be in 
  excellent agreement with the predictions provided in 
  Section~\ref{sec:wishart} (smooth curves). Note that the
  marginalised distribution for the TE cross power spectrum more
  closely resembles a Gaussian distribution than do the $\Gamma$
  distributions of the auto power spectra. 
  \label{fig:dist1}}
\end{figure*}

The marginal distributions of $\hat{C}_7^{\rm TT}$, $\hat{C}_7^{\rm
EE}$, $\hat{C}_7^{\rm TE}$ and $\hat{C}_7^{\rm BB}$ are plotted in
Fig.~\ref{fig:dist1} for the best-fit cosmological model of the WMAP
year-1 data \citep{spergel03} where we have included a B-mode
polarisation component with an input tensor-to-scalar ratio of $T/S = 0.05$. 
These distributions were calculated from all-sky realisations of
cosmological power spectra calculated using the CMBFAST 
\citep{seljak96} and HEALPIX \citep{gorski05} packages. 
For comparison we plot the distributions predicted by 
equation (\ref{eq:fCl_gamma}) for the auto power spectra and
equation (\ref{eq:C_TE_marg}) for the TE cross power spectrum, which show
excellent agreement with the simulated data.

For the Wishart distribution, if we define the data vectors of interest by
\be
  {\bf \hat{X}}_C = \left(\begin{array}{c} 
                           \hat{C}_l^{\rm TT} \\
                           \hat{C}_l^{\rm TE} \\
                           \hat{C}_l^{\rm EE} \end{array} \right),
  \,\,\,\,\,
  {\bf X}_C = \left(\begin{array}{c} 
                           C_l^{\rm TT} \\
                           C_l^{\rm TE} \\
                           C_l^{\rm EE} \end{array} \right),
  \label{eq:Xdef}
\ee
then the covariance matrix for ${\bf \hat{X}}_C$ is given by
\be
  {\bf Y} = \frac{1}{\nu}\left(\begin{array}{ccc} 
    2(C_l^{\rm TT})^2 & 
    2C_l^{\rm TT}C_l^{\rm TE} & 
    2(C_l^{\rm TE})^2 \\

    2C_l^{\rm TT}C_l^{\rm TE} & 
    C_l^{\rm TT}C_l^{\rm EE}+(C_l^{\rm TE})^2 & 
    2C_l^{\rm TE}C_l^{\rm EE} \\

    2(C_l^{\rm TE})^2 & 
    2C_l^{\rm TE}C_l^{\rm EE} & 
    2(C_l^{\rm EE})^2

    \end{array}\right).
  \label{eq:wishart_cov}
\ee
As expected, the variance for the auto-power spectra matches that of
the $\Gamma$ distribution discussed in equation (\ref{eq:fCl_gamma}).
However, the variance of the distribution in $\hat{C}_l^{\rm TE}$ has
a different form, reflecting the change in marginalised distribution
(equation (\ref{eq:C_TE_marg}) rather than a $\Gamma$ distribution).

In the limit $\nu\to\infty$, the Wishart distribution tends towards a
multi-variate Gaussian form, with the same covariance matrix. It is
worth noting that the matrix ${\bf Y}$ is also the curvature matrix
around the distribution maximum, which will become important in
Sections~\ref{sec:posterior_fit_single}
and~\ref{sec:posterior_fit_multi}.

\section{Complications for more realistic surveys}
\label{sec:complications}

As discussed in the previous Section, the posterior distribution for a
noise-less all-sky survey does not have a multi-variate Gaussian form
in the auto- and cross- power spectra at low $\nu$. The situation is
more complicated for realistic data that includes effects such as
noise, beam uncertainties, calibration errors and limited sky
coverage. Such effects are often dealt with by modifying the posterior
distribution to match that found from detailed simulations of the
particular experiment being considered (e.g. \citealt{hivon02}). This
modified distribution can, in turn, be fitted by simple functional
forms, thus allowing rapid calculation for any given cosmological
model (e.g. \citealt{bond00}). In this Section we briefly examine the
effect of real-world complications on the posterior, which provide
the motivation for our consideration of possible fits for the
posterior distribution in Section~\ref{sec:fits}.

\subsection{Partial sky coverage}
\label{sec:skycoverage}

First we simplify the analysis by ignoring noise and just considering
the effect of partial sky coverage. Here, the spherical harmonic
coefficients of the cut sky $\bar{a}_{lm}$ are related to the true
spherical harmonic coefficients by
\be
   \bar{a}_{lm} = \sum_{l'm'} K_{lm\,l'm'} a_{l'm'},
   \label{eq:alm_coupling}
\ee
where $ K_{lm\,l'm'}$ is a kernel which describes coupling between
modes introduced by the non-uniform sky weighting
\citep{hivon02,kogut03}. It is informative to consider why the Wishart
distribution is not valid for these modes. To see this, we focus on
the TT power spectrum. For the cut sky coefficients for a particular
mode, the estimator, $\hat{C}_l^{\rm TT}$, equation (\ref{eq:hatcl}),
now sums over a linear combination of the $a_{lm}$. We can work around
correlations between the $\bar{a}_{lm}$ by defining uncorrelated
combinations (which must be independent because they also form a
multi-variate Gaussian distribution). However, the distribution will
consist of variables with differing variance, and therefore does not
lead to a Wishart distribution. Additionally, modes at different $l$
will be correlated, again causing deviations from the analysis in
Section~\ref{sec:wishart}.

The introduction of any sky-cut to a CMB dataset renders the coupling
kernel, $K$ of equation (\ref{eq:alm_coupling}), singular, and thus
prohibits the use of equations (\ref{eq:hatcl}) \&
(\ref{eq:alm_coupling}) to estimate the underlying power
spectra. Various methods have therefore been developed for
estimating CMB power spectra from cut-sky datasets, the most prominent
of which are the pseudo-$C_l$ (PCL, \citealt{hivon02}) and quadratic
maximum likelihood (QML, \citealt{tegmark97}, \citealt{bond98}) methods.

There are two regimes that we can easily analyse - given a severe sky
cut, both the PCL and QML estimators will modify the likelihood
distribution from the full-sky Wishart distribution of
Section~\ref{sec:wishart}. However, on small scales, we can still
apply the central limit theorem to show that the likelihood of the
polarisation and temperature power spectra has a multi-variate
Gaussian distribution for both estimators. The covariance matrix of
this Gaussian distribution will change as a severe sky-cut gives rise
to the loss of information -- some linear combinations of the full-sky
${a}_{lm}$ will define modes that lie within the cut sky and these
modes cannot be recovered from the observed $\bar{a}_{lm}$. The net
result on the likelihood distribution is a reduction in the number of
degrees of freedom of the system. Consider, for example, the PCL
estimator for the temperature power spectrum defined by
\be
  \hat{C}_l = \sum_{l'} M^{-1}_{ll'} \widetilde{C}_{l'}, 
\ee 
where $\widetilde{C}_l = \frac{1}{2l+1} |\bar{a}^T_{lm}|^2$ is the
pseudo-power spectrum measured directly on the cut sky, and $M_{ll'}$
is the $C_l$ coupling matrix defined by,
\be 
M_{ll'} = \frac{(2l'+1)}{4\pi} \sum_{l''}{\mathcal W}_{l''} 
\left( \begin{array}{ccc} l & l' & l'' \\
0 & 0 & 0 \end{array} \right)^2.  
\ee 
Here, ${\mathcal W}_l$ is the power spectrum of the weight function,
$W(\Omega)$ applied to the data (e.g in the simplest case, $W=1$ for
observed pixels; $W=0$ otherwise) and the term in parentheses is the
Wigner 3-j symbol. On small scales, the recovered power spectrum
$\hat{C}_l$ will still have a $\Gamma$ distribution but with the
number of degrees of freedom reduced from $\nu = 2l + 1$ to
\citep{hivon02},
\be 
  \nu^{\rm eff} = (2l+1)f_{\rm sky}\frac{w_2^2}{w_4}, 
\ee 
where $w_i$ is the i$^{th}$ moment of the weighting scheme employed
and $f_{\rm sky}$ is the fraction of the sky having non-zero
weighting. If we ignore the effect of the weighting scheme, we see
that the covariances of the power spectra will be the standard all-sky
covariances divided by $f_{\rm sky}$ -- on small scales the number of
pairs of data points simply scales as the fraction of sky covered.

The other regime that can be easily analysed is a modest sky cut on
large scales. In this regime, the QML estimator approximates the exact
estimator of equation (\ref{eq:hatcl}) with estimates of the full-sky
$a_{lm}$'s given by
\be
   a_{lm} = \sum_{l'm'} \bar{K}^{-1}_{lm\,l'm'} \bar{a}_{l'm'},
   \label{eq:alm_recover}
\ee 
where $\bar{K}$ is a non-singular matrix formed by truncating the
coupling kernel, $K$ of equation (\ref{eq:alm_coupling}) at finite
values of $l'$ and $m'$ \citep{efstathiou04a}. This truncation of the
coupling kernel is possible since, at low multipoles, the
$\bar{a}_{lm}$ will only be weakly correlated with any of the $a_{lm}$
which lie within the cut sky. \cite{efstathiou04b} has demonstrated, using
simulations, that the QML estimator recovers the true $C_l$, at low $l$,
almost exactly in the presence of a modest sky cut ($f_{\rm sky} =
0.83$). For these QML $\hat{C}_l$ estimates, data at different $l$ are
independent and the likelihood distribution remains a Wishart
distribution as described in Section~\ref{sec:wishart}. In addition,
the (co)variances of the QML estimates remain those of equation
(\ref{eq:wishart_cov}).

Clearly, we are going to be interested in both of these regimes, and
in analysing data in between. Note that, in this latter regime, the
PCL estimator is known to be sub-optimal and the variances of PCL
estimates would be significantly increased by estimator-induced
variances.

The estimation of $\hat{C}_l$ is obviously coupled with the posterior
surface that should be assumed for any model. For instance a ``bias''
when a Gaussian posterior is assumed is removed when assuming the
correct shape of surface. The issue of whether different analysis
methods introduce further Gaussian or non-Gaussian noise is more
important for our present work.

\subsection{Including noise and beam smearing}
\label{sec:noise}

Here, we consider an all-sky CMB survey, with additive noise, $N_l$
and symmetric Gaussian beam smoothing, $B_l$. A single (independent)
auto-power spectrum measurement can now be written as $D_l=C_l
+N_l/B_l^2$, where we have dropped the explicit dependence on $TT$,
$EE$ or $BB$ -- when we do this, the formulae are valid for any of
these three auto-power spectra. Note that $D_l$ represents measured
spectra which have been corrected for the effect of the beam, $B_l$
but have not been corrected for the noise bias, $N_l$.  

At this point we need to make a distinction between isotropic and
anisotropic noise. For isotropic and uncorrelated Gaussian distributed
pixel noise, the $a_{lm}$ will be independent for different
$l,m$. However, this is not the case for anisotropic noise, where the
noise level changes from pixel-to-pixel. In this latter situation, we
have, in effect, a noise window function that will induce correlations
between the $a_{lm}$. In the former case of isotropic pixel noise,
$N_l = {\rm const}$ and our data vector (of which we are calculating
the power) is still expected to have a Gaussian distribution. In this
case, the marginalised distribution of $\hat{D}_l$ still has a
$\Gamma$ distribution, and the posterior is altered from equation
(\ref{eq:Lallsky}), becoming \citep{bond00} \be -2 \ln
f(D_l|\hat{D}_l) = (2l+1) \left[\ln(C_l+N_l/B_l^2) +
\frac{\hat{D}_l}{C_l+N_l/B_l^2} \right].
   \label{eq:L_cmb_with_err}      
\ee

The curvature around the posterior maximum is
\be                                                                             
   -\frac{d^2 f(D_l|\hat{D}_l)}{dC_l dC_{l'}} \propto                           
     (C_l+N_l/B_l^2)^{-2} \delta_{ll'},                                         
   \label{eq:fisher2}                                                           
\ee
and the error on our model power spectrum $C_l$ is proportional to
$(C_l+N_l/B_l^2)$.
                                                                                
In the case of a joint analysis of temperature and polarisation data,
for isotropic noise, the matrix ${\bf S}_l$, equation (\ref{eq:Sl}),
will still obey the Wishart distribution of equation
(\ref{eq:wishart}), but with a revised ${\bf W}_l$ matrix given by
${\bf W}'_l = {\bf V}'_l / (2l + 1)$ where ${\bf V}'_l$ is now given
by
\be                                                                             
   {\bf V}'_l = \left(\begin{array}{ccc}                                        
     C_l^{\rm TT} + N_l^{\rm T}/(B_l^{\rm T})^2 & C_l^{\rm TE} & 0 \\
     C_l^{\rm TE} & C_l^{\rm EE} + N_l^{\rm P}/(B_l^{\rm P})^2 & 0 \\
     0 & 0 & C_l^{\rm BB} + N_l^{\rm P}/(B_l^{\rm P})^2 
    \end{array}\right).
   \label{eq:vl_rev}                                                            
\ee
In equation (\ref{eq:vl_rev}), we have now allowed for different beam
widths ($B_l^{\rm T}$ \& $B_l^{\rm P}$) and noise levels ($N_l^{\rm
  T}$ \& $N_l^{\rm EE} = N_l^{\rm BB} \equiv N_l^{\rm P}$) for the
temperature and polarisation data. Note that for uncorrelated Gaussian
distributed noise, the noise cross power spectrum, $N_l^{TE} = 0$. The
inclusion of noise and beam smoothing will modify the covariance
matrix of the power spectra measurements, equation
(\ref{eq:wishart_cov}), which becomes \citep{eisenstein99}
\be
  {\bf Y}' = \frac{1}{\nu}\left(\begin{array}{ccc} 
    2(C_l^{\rm TT} + N^{\rm T}_l / (B^{\rm T}_l)^2 )^2 & 
    2C_l^{\rm TE}(C_l^{\rm TT} + N^{\rm T}_l / (B^{\rm T}_l)^2) & 
    2(C_l^{\rm TE})^2 \\

    2C_l^{\rm TE}(C_l^{\rm TT} + N^{\rm T}_l / (B^{\rm T}_l)^2) & 
    (C_l^{\rm TE})^2+(C_l^{\rm TT}+N^{\rm T}_l / (B^{\rm T}_l)^2) 
    (C_l^{\rm EE}+N^{\rm P}_l / (B^{\rm P}_l)^2) & 
    2C_l^{\rm TE} ( C_l^{\rm EE} + N^{\rm P}_l / (B^{\rm P}_l)^2) \\

    2(C_l^{\rm TE})^2 & 
    2C_l^{\rm TE} ( C_l^{\rm EE} + N^{\rm P}_l / (B^{\rm P}_l)^2) &
    2(C_l^{\rm EE} + N^{\rm P}_l / (B^{\rm P}_l)^2 )^2 \\

    \end{array}\right).
  \label{eq:wishart_cov_new}
\ee
                                                             
These expressions, which are for the case of uniform uncorrelated
pixel noise, are most relevant for satellite experiments where the
pixel noise covariance matrix is near-diagonal. However, the planned
scanning strategy of the Planck satellite mission will result in a
much larger integration time near the ecliptic poles, leading to an
anisotropic noise map. This experiment, and ground-based experiments
with more complicated noise properties, will require simulations to
accurately quantify the posterior distribution for the system. Such
considerations motivate the consideration of fits to the posterior
distribution considered in the next section.

\section{Fitting to the posterior}
\label{sec:fits}

The complications discussed in the previous section for realistic
data, and the deviations from Gaussian behaviour of the posterior
discussed in Section~\ref{sec:likepost-all-sky}, can be modelled using
a fit to the posterior surface. Such fitting functions can also allow
the posterior surface to be approximated without requiring the
inversion of a covariance matrix for every model tested -- the effect
of a varying covariance matrix can be absorbed into the shape of the
function. In Section~\ref{sec:posterior_fit_single} we consider a
number of possible forms for the posterior fit to a single auto-power
spectrum. The extension to include polarisation data is considered in
Section~\ref{sec:posterior_fit_multi}. A good choice of fitting
function should be able to interpolate between the posterior
distribution in the limit of an all-sky survey at low-$l$ and in the
limit of a multi-variate Gaussian distribution at high-$l$. In order
to compare the suitability of different fitting functions, we
therefore choose to consider their ability to match the true
distributions in these situations. As the inclusion of isotropic
and uncorrelated Gaussian distributed pixel noise does not change the
shape of the posterior distributions in either limit, we can ignore
its contribution without loss of generality -- the effect of isotropic
noise can be trivially included in both the fits and in the limiting
situations that we are testing against.

\subsection{Single mode auto-power spectra}
\label{sec:posterior_fit_single}

We now consider a number of possible fitting functions for the
posterior distribution for a single mode of an auto-power spectrum. As
in the previous section we will drop the explicit reference to a
particular auto-power spectrum, as the formulae and concepts are valid
for TT, EE or BB power spectra. We quantify the shapes of different
fits using an expansion of the posterior around the maximum $C_l =
(1+\epsilon)\hat{C}_l$ (a variation of the method used in
\citealt{verde03}).

For an all-sky, no-noise survey, equation (\ref{eq:Lallsky}) can be
expanded to give
\be
  -2\ln f(C_l|\hat{C}_l) = \nu\left[\frac{\epsilon^2}{2}
                          -\frac{2\epsilon^3}{3}+O(\epsilon^4)\right],
  \label{eq:L_epsilon_true}
\ee
ignoring an irrelevant additive constant. Note that this is not equal
to equation~9 in \citet{verde03} because of the different expansion
adopted (we expand around $\hat{C}_l$ rather than $C_l$). At first
glance, it appears that changing the value of $\nu\equiv(2l+1)$ does
not change the shape of the surface around the maximum as it affects
all order terms equally. However, as $\nu$ increases, the posterior at
a fixed value of $\epsilon$ increases. We therefore see that, although
the overall shape does not change, the range of parameters with $-2\ln
f(C_l|\hat{C}_l)$ below a fixed value reduces in size and the term of
order $\epsilon^2$ dominates the behaviour. It is in this way that the
distribution tends to a Gaussian as $\nu\to\infty$.

For an all-sky survey, the curvature around the posterior maximum is
$2(\hat{C}_l)^2/\nu$. Given a distribution $f(C_l|\hat{C}_l)$, then
the term of order $\epsilon^2$, where $C_l = (1+\epsilon)\hat{C}_l$,
should equal $\nu \epsilon^2/2$ in order to match this curvature. The
distributions that we now consider as approximations to the posterior
surface have all been normalised to match this behaviour to order
$\epsilon^2$. Note that these distributions depend on the model power
and therefore intrinsically allow for the model dependence of the
variances of the measured power spectra. In this case, a fixed
covariance matrix should be used with these posterior fits. An
alternative approach would be to recalculate the variance for each
model and use an altered posterior shape
\citep{efstathiou04b,challinor05,brown05}.

\begin{enumerate}
\renewcommand{\theenumi}{(\arabic{enumi})}

\item First, we consider a $\Gamma$ distribution for $C_l$ with
degrees-of-freedom equal to $\mu$, thereby matching the shape of the
likelihood. We have to be slightly careful as the distribution is
usually defined in terms of the mean rather than the maximum. In terms
of the maximum, $\hat{C}_l$, assuming a $\Gamma$ distribution gives
\be f(C_l|\hat{C}_l) \propto (C_l)^{\mu/2-1} \exp\left[-\frac{(\mu-2)
C_l}{2\hat{C}_l}\right], \ee where we have an extra $(\mu-2)/\mu$ term
in the exponent compared with equation (\ref{eq:fCl_gamma}).

This leads to an expansion in $\epsilon$
\be
  -2\ln f(C_l|\hat{C}_l)
    \propto (\mu-2)\left[-\ln (C_l/\hat{C}_l)
                           + \frac{C_l-\hat{C}_l}{\hat{C}_l}\right]
    = (\mu-2)\left[\frac{\epsilon^2}{2}-\frac{\epsilon^3}{3}
                         +O(\epsilon^4)\right].
  \label{eq:chisquared}
\ee
This distribution is included to emphasise the fact that the posterior does not
have a the same form as the likelihood -- this can easily be
seen by comparing equations (\ref{eq:L_epsilon_true}) and
(\ref{eq:chisquared}).

\item Next, we consider a Gaussian distribution in $C_l$ with fixed
variance -- chosen to match the curvature around the distribution
maximum
\be
  -2\ln f(C_l|\hat{C}_l) 
    \propto \frac{\nu}{2(\hat{C}_l)^2}
    \left[(C_l-\hat{C}_l)\right]^2
     = \nu\left[\frac{\epsilon^2}{2}\right].
  \label{eq:Gauss_fix}
\ee
The mean and maximum of this fit coincide at $\hat{C}_l$. Most of the
distributions that we now consider result from replacing $C_l$ and
$\hat{C}_l$ in the right hand side of this expression with functions
$g(C_l)$ and $g(\hat{C}_l)$.

\item For example, setting $g(C_l)=(\hat{C}_l)^2/C_l$ gives 
\be
  -2\ln f(C_l|\hat{C}_l)
    \propto \frac{\nu}{2(\hat{C}_l)^2}
    \left[\frac{\hat{C}_l}{C_l}(C_l-\hat{C}_l)\right]^2
    = \nu\left[\frac{\epsilon^2}{2}-\epsilon^3
                         +O(\epsilon^4)\right].
  \label{eq:Gauss_vary}
\ee
This distribution could also have been derived from equation
(\ref{eq:Gauss_fix}), by including a factor $(C_l/\hat{C}_l)^2$ to
make the variance a function of the model. Note that this distribution
is not a Gaussian distribution in $C_l$ with covariances proportional
to $(C_l)^2$ -- this would require an extra term in the posterior from
the effect on the determinant of the covariance matrix. Additionally,
this distribution is not the same as a Gaussian in
$(\hat{C}_l)^2/C_l$, which would require the inclusion of a Jacobian
from the change of variables.

\item In a recent paper \citet{smith06} have suggested a form for the
posterior with $g(C_l)=3(\hat{C}_l)^{4/3}(C_l)^{-1/3}$, which gives
\be
  -2\ln f(C_l|\hat{C}_l)
    \propto \frac{\nu}{2(\hat{C}_l)^2}
    9\left[\hat{C}_l - (\hat{C}_l)^{4/3}(C_l)^{-1/3}\right]^2
    = \nu\left[\frac{\epsilon^2}{2}
               -\frac{2\epsilon^3}{3}+O(\epsilon^4)\right].
  \label{eq:Cthird}
\ee
The $(C_l)^{-1/3}$ formula was derived in \citet{smith06} from the
third derivative of the posterior of an all-sky survey, and can be
seen to recover the correct behaviour to order $\epsilon^3$, matching
equation (\ref{eq:L_epsilon_true}).

\item The standard log-normal distribution can be derived by setting
$g(C_l)=\hat{C}_l\ln C_l$, to give
\be
  -2\ln f(C_l|\hat{C}_l)
    \propto \frac{\nu}{2(\hat{C}_l)^2}
    \left[\hat{C}_l\ln(C_l/\hat{C}_l)\right]^2
    = \nu\left[\frac{\epsilon^2}{2}
               -\frac{\epsilon^3}{2}+O(\epsilon^4)\right].
  \label{eq:LN_fix}
\ee

\item In analogy with equations (\ref{eq:Gauss_fix}) \& (\ref{eq:Gauss_vary}),
we can consider $g(C_l)=(\hat{C}_l)^2\ln(C_l)/C_l$ (non-standard log-normal).
\be
  -2\ln f(C_l|\hat{C}_l)
    \propto \frac{\nu}{2(\hat{C}_l)^2}
    \left[\frac{(\hat{C}_l)^2}{C_l}\ln(C_l/\hat{C}_l)\right]^2
    = \nu\left[\frac{\epsilon^2}{2}
               -\frac{3\epsilon^3}{2}+O(\epsilon^4)\right].
  \label{eq:LN_vary}
\ee
This is the distribution obtained by setting the variance to be a
function of the model to be tested in the standard log-normal
distribution, equation (\ref{eq:LN_fix}).

\item It is also possible to consider the offset log-normal
distribution, where $g(C_l)=\hat{C}_l(1+a)\ln(C_l+a\hat{C}_l)$
\be
  -2\ln f(C_l|\hat{C}_l)
    \propto \frac{\nu}{2(\hat{C}_l)^2}
    \left[\hat{C}_l(1+a)\ln(\frac{C_l+a\hat{C}_l}
                           {\hat{C}_l+a\hat{C}_l})\right]^2
    = \nu\left[\frac{\epsilon^2}{2}
               -\frac{1}{2(1+a)}\epsilon^3+O(\epsilon^4)\right].
  \label{eq:expansion_offset_LN}
\ee
Setting $a=-1/4$ matches the all-sky no noise behaviour to order
$\epsilon^3$. As $a\to\infty$, this distribution tends towards
Gaussian form.

\item Because all of our definitions of $g(C_l)$ require the same
curvature matrix to match the $\epsilon^2$ behaviour of the true
distribution, we could also consider a combination of 2 or more of
them. For example, setting $g(C_l)=aC_l+(1-a)\hat{C}_l\ln C_l$ gives
\be
  -2\ln f(C_l|\hat{C}_l)
    \propto \frac{\nu}{2(\hat{C}_l)^2}
    \left[a(C_l-\hat{C}_l) + (1-a)\hat{C}_l\ln(C_l/\hat{C}_l)\right]^2
    = \nu\left[\frac{\epsilon^2}{2}+\frac{a-1}{2}\epsilon^3
               +O(\epsilon^4)\right].
  \label{eq:expansion_summed_LN}
\ee
As with the offset log-normal distribution, we can set the free
parameter $a=-1/3$ to match the behaviour of the true distribution to
order $\epsilon^3$. For $a\to1$, the distribution obviously tends
towards a Gaussian form. This distribution, which we suggest calling a
summed log-normal distribution, was used by \citet{percival05} to
model the large-scale structure power spectrum from the 2dF galaxy
redshift survey.

\item An alternative procedure, adopted by \citet{verde03}, is to
combine different posteriors after calculation. \citet{verde03}
considered combining the posterior ${\cal P}_{\rm 1}$ of
equation (\ref{eq:Gauss_vary}) and the posterior ${\cal P}_{\rm 2}$ of
equation (\ref{eq:LN_fix}). Matching the all-sky
no-noise posterior shape of equation (\ref{eq:L_epsilon_true}) requires
\be
  -2\ln f(C_l|\hat{C}_l)
    \propto \frac{1}{3}{\cal P}_{\rm 1} 
          + \frac{2}{3}{\cal P}_{\rm 2}
    = \nu\left[\frac{\epsilon^2}{2}-\frac{2\epsilon^3}{3}
               +O(\epsilon^4)\right].
\ee

\end{enumerate}

We have presented a variety of possible fits to the posterior surface
in order to highlight that, even for a single auto-power spectrum, the
optimal choice is by no means certain, and will depend on the
experiment being analysed.

\subsection{Single mode combined temperature-polarisation spectra}
\label{sec:posterior_fit_multi}

We now extend our consideration of fitting formulae for the posterior
presented in Section~\ref{sec:posterior_fit_single} to include
combined temperature and polarisation data. In this section we only
consider a single-mode, and our vector of model power spectra is
${\bf X}_C$, with observed value ${\bf \hat{X}}_C$, defined in
equation (\ref{eq:Xdef}).

For an all-sky no-noise survey, the marginalised likelihood
distribution for the temperature-polarisation cross-power spectrum
$\hat{C}_l^{\rm TE}$ was given in equation (\ref{eq:C_TE_marg}), and has a
form that is different from that of the auto-power spectra. It is
therefore clear that the posterior predicted for $C_l^{\rm TE}$ will
have a different shape from that of equation (\ref{eq:Cl_posterior}). In
fact, the maximum in the posterior of the marginalised distribution no
longer occurs at $C_l^{\rm TE}=\hat{C}_l^{\rm TE}$, so we cannot
expand the marginalised distribution around the maximum. However, the
general Wishart distribution presented in Section~\ref{sec:wishart}
does predict a maximum in the posterior at ${\bf X}_C={\bf
  \hat{X}}_C$, so we can expand around this point. We can write the
posterior distribution as
\be
  f({\bf X}_C|{\bf \hat{X}}_C) \propto
    \frac{1}{[C_l^{\rm TT}C_l^{\rm EE}-(C_l^{\rm TE})^2]^{\nu/2}}
    \exp\left[-\frac{\nu}{2}\left(
    \frac{C_l^{\rm TT}\hat{C}_l^{\rm EE}
          +\hat{C}_l^{\rm TT}C_l^{\rm EE}-2\hat{C}_l^{\rm TE}C_l^{\rm TE}}
         {C_l^{\rm TT}C_l^{\rm EE}-(C_l^{\rm TE})^2}\right)\right].
\ee
The curvature matrix around the maximum can be calculated
from the second derivatives of this distribution giving
\be
  {\bf Y}_l^{-1} = \frac{\nu}{2
    [\hat{C}_l^{\rm TT}\hat{C}_l^{\rm EE}-(\hat{C}_l^{\rm TE})^2]^2}
  \left(\begin{array}{ccc} 
    (\hat{C}_l^{\rm EE})^2 & 
    -2\hat{C}_l^{\rm EE}\hat{C}_l^{\rm TE} & 
    (\hat{C}_l^{\rm TE})^2 \\
    -2\hat{C}_l^{\rm EE}\hat{C}_l^{\rm TE} & 
    2[\hat{C}_l^{\rm TT}\hat{C}_l^{\rm EE}+(\hat{C}_l^{\rm TE})^2] & 
    -2\hat{C}_l^{\rm TT}\hat{C}_l^{\rm TE} \\
    (\hat{C}_l^{\rm TE})^2 & 
    -2\hat{C}_l^{\rm TT}\hat{C}_l^{\rm TE} &
    (\hat{C}_l^{\rm TT})^2
    \end{array}\right),
  \label{eq:curvature_matrix}
\ee 
which is the inverse of the covariance matrix given in
equation (\ref{eq:wishart_cov}) at ${\bf X}_C={\bf \hat{X}}_C$.

To compare fitting functions with this distribution, we will adopt the
philosophy used in Section~\ref{sec:posterior_fit_single} for
independent auto-power spectra, and expand around the maximum. The
equations are simplified if we define
\be
  r       = \frac{C_l^{\rm TE}}
                 {\sqrt{\hat{C}_l^{\rm TT}\hat{C}_l^{\rm EE}}},\,\,
  \hat{r} = \frac{\hat{C}_l^{\rm TE}}
                 {\sqrt{\hat{C}_l^{\rm TT}\hat{C}_l^{\rm EE}}}.
\ee

First we expand the Wishart distribution in the direction of $C_l^{\rm
  TT}$ by fixing $C_l^{\rm TE}=\hat{C}_l^{\rm TE}$ \& $C_l^{\rm
  EE}=\hat{C}_l^{\rm EE}$, and expanding in $C_l^{\rm
  TT}=(1+\epsilon)\hat{C}_l^{\rm TT}$. In this direction,
\be
  -2\ln f(C_l^{\rm TT}|\hat{C}_l^{\rm TT}) \propto 
    \nu\left[\frac{\epsilon^2}{2}
             +\frac{2}{3(\hat{r}^2-1)}\epsilon^3
             +O(\epsilon^4)\right].
  \label{eq:wishart_rexpansion1}
\ee
In the limit as $\hat{r}\to0$, this tends towards the distribution
given in equation (\ref{eq:L_epsilon_true}) as expected.  By symmetry,
expanding $C_l^{\rm EE}$ around $\hat{C}_l^{\rm EE}$ would give the same series
expansion (the auto power spectra predict the same posterior shape).

Expanding the Wishart distribution in the direction of $C_l^{\rm TE}$
by fixing $C_l^{\rm TT}=\hat{C}_l^{\rm TT}$ \& $C_l^{\rm
  EE}=\hat{C}_l^{\rm EE}$, and setting $C_l^{\rm
  TE}=(1+\epsilon)\hat{C}_l^{\rm TE}$ gives
\be
  -2\ln f(C_l^{\rm TE}|\hat{C}_l^{\rm TE}) \propto 
    \nu\left[\frac{\epsilon^2}{2}
             +\frac{2\hat{r}^2(\hat{r}^2+3)}
                   {3(\hat{r}^4-1)}\epsilon^3
             +O(\epsilon^4)\right].
  \label{eq:wishart_rexpansion2}
\ee
For $\hat{r}^2=0$, the distribution has a perfect Gaussian form.

We now consider fitting this distribution using multi-variate
extensions of the functions discussed in
Section~\ref{sec:posterior_fit_single} for auto-power spectra. The
expansions in the directions of the auto and cross-power spectra have
different shapes, so the {\em shape} of the fitted distribution for
$C_l^{\rm TE}$ must differ from that of $C_l^{\rm TT}$ and $C_l^{\rm
  EE}$. The offset log-normal distribution has the flexibility to
allow for this change in shape, and we will focus on the multi-variate
extension of this distribution. Formally, we will consider a
distribution in
\be
  {\bf Z}_C = \left(\begin{array}{c} 
    \hat{C}_l^{\rm TT}(1+a^{\rm TT})
        \ln(C_l^{\rm TT}+a^{\rm TT}\hat{C}_l^{\rm TT}) \\
    \hat{C}_l^{\rm TE}(1+a^{\rm TE})
        \ln(C_l^{\rm TE}+a^{\rm TE}\hat{C}_l^{\rm TE}) \\
    \hat{C}_l^{\rm EE}(1+a^{\rm EE})
        \ln(C_l^{\rm EE}+a^{\rm EE}\hat{C}_l^{\rm EE})
 \end{array}\right),
 \label{eq:offset_ln}
\ee
given by
\be
  -2\ln f({\bf X}_C|{\bf \hat X}_C) \propto 
    ({\bf Z}_C-{\bf \hat{Z}}_C)'{\bf Y}_l^{-1}
    ({\bf Z}_C-{\bf \hat{Z}}_C).
\ee

The series expansion of this multi-variate distribution along the
standard axes was given in equation (\ref{eq:expansion_offset_LN}). Matching
this expansion to equations (\ref{eq:wishart_rexpansion1})
\& (\ref{eq:wishart_rexpansion2}), gives
\be
  a^{\rm TT} = a^{\rm EE} = -\frac{1}{4}(1+3\hat{r}^2),\,\,\,\,
  a^{\rm TE} =
    -\frac{1}{2}\left(2+\frac{3(\hat{r}^4-1)}{2\hat{r}^2(\hat{r}^2+3)}\right).
\ee

\begin{figure*}
  \centering
  \resizebox{0.80\textwidth}{!}{\includegraphics{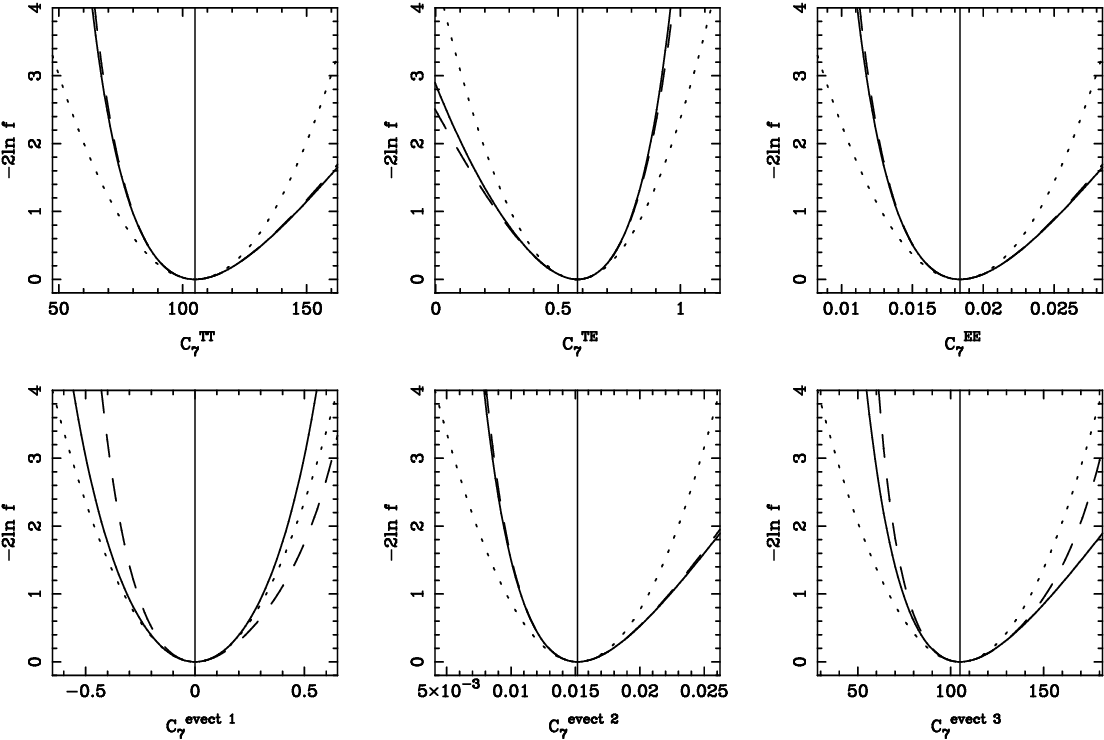}}

    \vspace{1cm}

  \resizebox{0.80\textwidth}{!}{\includegraphics{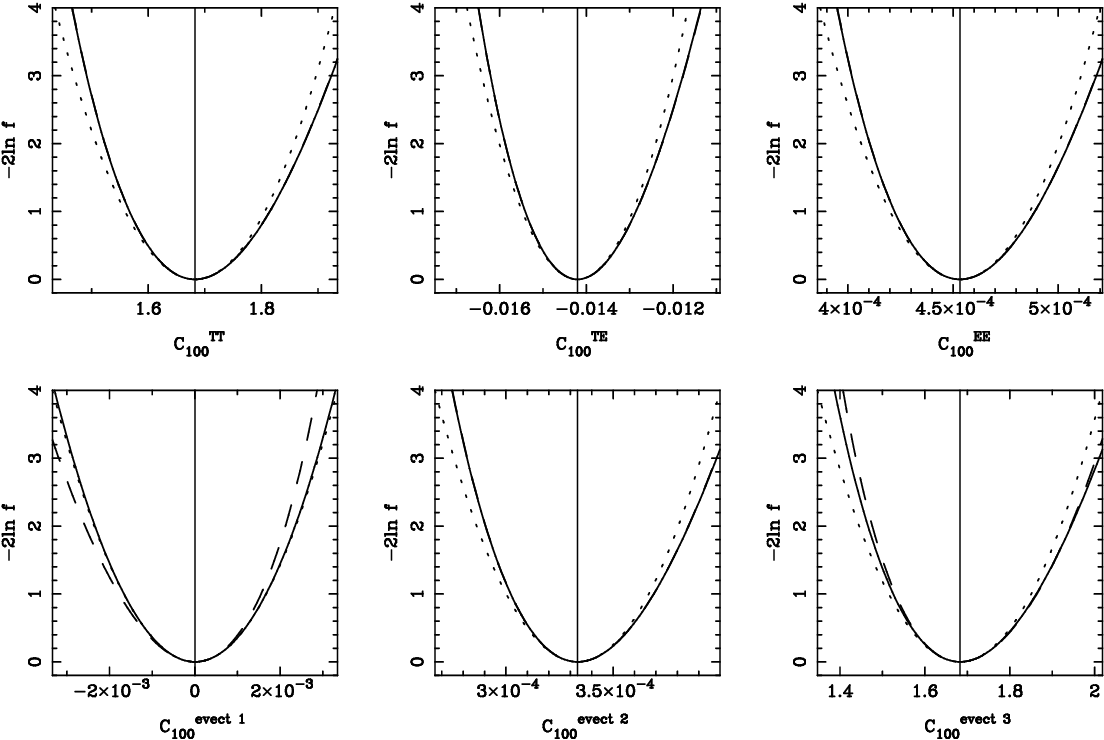}}
  \caption{Posterior distributions in different directions in
  parameter space calculated for the Wishart distribution (solid black
  line), a simple Gaussian fit to this distribution (dotted line), and
  a log-normal distribution matched to the Wishart distribution
  (dashed line). Plots are presented for $l=7$ (top two rows) and
  $l=100$ (bottom two rows). For each value of $l$, the upper row
  shows the distribution along the auto and cross-power spectra, while
  the lower row shows the distribution expanded along the eigenvectors
  of the covariance matrix. Cosmological parameters were fixed at the
  best-fit values of the 6-parameter model fitted to the CMB and
  2dFGRS data in \citet{sanchez06}.
    \label{fig:l7_l100_offset_LN}}
\end{figure*}

Note that the logarithmic terms in equation (\ref{eq:offset_ln}) can
be ill-defined for certain models. This is true already for the
offset-lognormal distribution commonly used to approximate the
$C_l^{TT}$ likelihood distribution -- in equation
(\ref{eq:offset_ln}), $a^{\rm TT}$ is a free parameter and can, in
principle, take negative values leading to an ill-defined likelihood
for particular models. However, letting $a \rightarrow \infty$ (for
+ve measured $C_l$) or $a \rightarrow -\infty$ (for -ve measured
$C_l$) will make the distributions in equation (\ref{eq:offset_ln})
tend to a Gaussian, and the fit is well-defined for all models in this
limit. In general, we're only concerned with the slightly non-Gaussian
regime where ill-defined terms do not arise except for very extreme
models. One should therefore consider the posterior distribution of
equation (\ref{eq:offset_ln}) as only being valid for a subset of
models - those models for which the fit is well defined.  All other
models are assumed to be highly unlikely. Note that this does not
prohibit the use of this function as a fitting function -- in fact, as
can be seen in Fig.~\ref{fig:l7_l100_offset_LN}, it matches the
all-sky likelihood well in the directions of the cross- and auto-power
spectra.

The distributions calculated adopting these parameters and the
curvature matrix of equation (\ref{eq:curvature_matrix}) are plotted
in Fig.~\ref{fig:l7_l100_offset_LN}, for a basic cosmological model
with parameters set at the best-fit values of the simplest 6-parameter
model that adequately fits the CMB and 2dFGRS data presented in
\citet{sanchez06}. We present distributions at two wavenumbers $l=7$,
and $l=100$. Fixing the distribution shape in the directions of the
power spectra leaves no free parameters in this simple fit as the
covariance matrix is fixed to match the curvature around the peak.  In
addition to matching the distributions along the power spectra
elements of ${\bf X}_C$, the distribution should match the posterior
distribution in arbitrary directions in parameter space.
In Fig.~\ref{fig:l7_l100_offset_LN} we therefore also plot the
distribution along the principal components of the covariance matrix
for these two values of $l$. As can be seen, fitting the distribution
along the power spectra does not constrain the distribution in other
directions with the same accuracy.  This is an inadequacy of our
assumption of replacing the power spectra in equation
(\ref{eq:Gauss_fix}) with a more general function, rather than a
problem in the accuracy of the function chosen. We can match the
function arbitrarily well in the directions of the power spectra, but
the discrepancy in other directions remains. For example, in addition
to using the offset log-normal distribution, we have also considered
simply defining new variables as the sum of powers of $(C_l^{\rm
XX}-\hat{C}_l^{\rm XX})$, which can be adjusted to match the
distribution to arbitrary order around the maximum. However, even in
this case, we find similar problems to those encountered using the
offset log-normal distribution. Had we fixed these simple fits in the
directions of the principal components, then we would have found
problems fitting along the directions of the power spectra. One can,
of course, envisage constructing more complicated fitting functions
which more accurately reproduce correlations between different power
spectra. Such functions would, however, lack the simplicity (and
therefore some of the appeal) of fits commonly used to model the
temperature power spectrum posterior.

\section{Marginalising over nuisance parameters} \label{sec:marginalise}

For real CMB datasets, one often needs to account for an uncertainty in
the calibration of the experiment. This uncertainty is usually
considered a nuisance parameter and is marginalised over. For simple
assumptions about the form of the underlying power spectrum posterior
and of the calibration error, we can perform this marginalisation
analytically. Here, we consider the effect of the posterior shape on
this marginalisation process. For this analysis, we focus on a single
auto-power spectrum. As in previous sections, we drop the explicit
dependence on $TT$, $EE$ or $BB$ when the formulae and derivation are
valid for any of these three auto-power spectra.

Consider an experiment where the observed data has a multiplicative
``calibration'' error, $b$ that is known to have a Gaussian
distribution ($\langle b\rangle=0$, $\langle
b^2\rangle=\sigma_b^2$). If we know the calibration error, then the
``true'' observed power spectrum value can be recovered, 
$(\hat{C}_l)_{\rm true}=(1+b)\hat{C}_l$. The posterior distribution of
$C_l$ is then given by
\be
  f(C_l|\hat{C}_l,\sigma_b) = \int db f(C_l,b|\hat{C}_l,\sigma_b)
    = \int db f(C_l|\hat{C}_l,b,\sigma_b)f(b|\sigma_b).
\ee
If $f(C_l|\hat{C}_l,b,\sigma_b)$ has a Gaussian form with variance $S$
then, 
\be
  f(C_l|\hat{C}_l,b,\sigma_b) = \frac{1}{\sqrt{2\pi S}} 
    \exp\left[-\frac{1}{2}(C_l-(1+b)\hat{C}_l)
              S^{-1}(C_l-(1+b)\hat{C}_l)\right].
\ee
This marginalisation can be reduced by ``completing the square''
\citep{bridle02} to give
\be
  f(C_l|\hat{C}_l,\sigma_b) = (1+C_lS^{-1}C_l\sigma_b^2)^{-1/2}
      f(C_l|\hat{C}_l,b',\sigma_b),
  \label{eq:gauss_marg1}
\ee
where 
\be
  b'=\frac{C_lS^{-1}\hat{C}_l
           -\hat{C}_lS^{-1}\hat{C}_l}
	  { \hat{C}_lS^{-1}\hat{C}_l+\sigma_b^{-2}},
  \label{eq:gauss_marg2}
\ee
 is the value of the calibration that maximises
$f(C_l,b|\hat{C}_l,\sigma_b)$. The offset term
$(1+C_lS^{-1}C_l\sigma_b^2)^{-1/2}$ in equation (\ref{eq:gauss_marg1})
arises because the variance $S$ is independent of the calibration
error. 

The procedure for analytic marginalisation is, however, dependent on the
posterior distribution. For an auto-power spectrum of an all-sky
no-noise survey, the joint distribution of $C_l$ and $b$ is given by
\be
  f(C_l,b|\hat{C}_l,\sigma_b) \propto (C_l)^{-\nu/2}
    \exp\left[-\frac{\nu(1+b)\hat{C}_l}{2C_l}\right]\,
    \frac{1}{\sqrt{2\pi}\sigma_b}\exp\left[-\frac{b^2}{2\sigma_b^2}\right].
\ee
As in Section~\ref{sec:fits}, we have ignored a contribution from
uncorrelated Gaussian pixel noise, which can be easily included as it
does not affect the shape of the posterior distribution. Completing
the square (as for the Gaussian case above) gives the marginalised
posterior,
\be
  f(C_l|\hat{C}_l,\sigma_b) \propto (C_l)^{-\nu/2}
    \exp\left[-\frac{\nu(1+2b')\hat{C}_l}{2C_l}\right],
  \label{eq:anal_marg1}
\ee
where
\be
  b'=-\frac{\nu\hat{C}_l\sigma_b^2}{2C_l},
  \label{eq:anal_marg2}
\ee
is the value of the calibration error that maximises the distribution
$f(C_l,b|\hat{C}_l,\sigma_b)$. As can be seen, analytic
marginalisation is also trivial when the exact posterior distribution
for an all-sky survey is used, rather than a Gaussian. Because of the
skewness of this distribution, the result is offset -- the value of
the calibration error needed to mimic the effect of a full
marginalisation is twice that which maximises the likelihood. This
demonstrates the difference between marginalisation and simply taking
the likelihood maximum.

Our decision to adopt a Gaussian distribution for the calibration
error for a single auto-power spectrum measurement was reasonably
arbitrary, and we could instead, have considered a calibration error
that is Gaussian in the pixel values. When considering the joint
likelihood of temperature and polarisation data, the nature of the
calibration error becomes increasingly important. For example,
assuming independent Gaussian distributed temperature ($b^{\rm T}$)
and polarisation ($b^{\rm P}$) calibration errors would give a
calibration error on the TE power spectrum with a distribution based
on a modified Bessel function -- the distribution of $[(1+b^{\rm
T})(1+b^{\rm P})]^{1/2}$. Obviously, this would complicate the
procedure for analytic marginalisation. Additionally, from a single
experiment, the temperature and polarisation calibration errors will
be highly correlated, which will further increase the
complications. Given these issues, there is clearly little to be
gained from working through a derivation of analytic marginalisation
for combined temperature and polarisation power spectra. For a given
experiment, it seems clear that numerical simulations would be
required to quantify the effect on the posterior.

\section{Discussion}
\label{sec:discuss}
The ultimate goal of CMB experiments is to constrain cosmological
models by comparison with theory. In the currently favoured
inflationary based cosmological model, the temperature and
polarisation fluctuations in the CMB are expected to be isotropic and
approximately Gaussian distributed \citep{liddle00}. In this case,
the statistical properties of the CMB can be described completely by
the auto- and cross-power spectra of the temperature and polarisation
fields. This data compression greatly speeds up the process of
comparing with large numbers of theoretical models. In this paper, we
have reviewed the mathematical foundation for this comparison. In a
Bayesian analysis the posterior, which determines the model
constraints, is directly related to the likelihood of a set of data
given a particular model; it is therefore important to characterise
the likelihood for a given experiment.

For an all-sky survey with uncorrelated Gaussian distributed pixel
noise, we have shown that the joint likelihood of the four CMB power
spectra is given by a Wishart distribution, a distribution commonly
encountered when calculating covariance matrices from Gaussian
distributed data. This distribution, which can most easily be written
in terms of matrices of the data and model power spectra, provides the
likelihood of the measured power spectra including the constraint of
positive definiteness. The shape of the likelihood is significantly
different from a multi-variate Gaussian at low order multipoles,
although it tends towards a multi-variate Gaussian form at high
multipoles. The Wishart distribution can be integrated to give
marginal distributions for the individual auto- and cross-power
spectra. For the auto-power spectra, these marginalised distributions
reduce to the well known $\Gamma$ functions. For the TE cross-power
spectrum, the marginalised distribution is more complicated, but can
be calculated. We find that the resulting distribution for TE is
significantly different to the $\Gamma$ distributions of the auto
power spectra and is closer to (although still differs from) a
Gaussian.  We have compared the marginalised distributions with those
empirically determined from simulated data, finding excellent
agreement. Realistically, CMB observations that include polarisation
measurements will simultaneously provide constraints on all of the
auto- and cross-power spectra. Consequently, the marginal
distributions are probably only of academic interest, and we need to
consider the combined distribution of all of the different power
spectra.

Given the complications of noise and limited sky coverage in real CMB
data (discussed in Section~\ref{sec:complications}), the distribution
of measured power spectra will, in general, deviate from a Wishart
distribution to some degree. However, for a moderate sky cut, it is
possible to recover the true temperature auto-power spectrum on large
scales. In the case of polarisation data, large uncertainties in the
level of polarised foregrounds are currently a limiting factor for CMB
experiments. If our understanding of such foregrounds improves
sufficiently, then it may be possible to recover the full-sky versions
of all four CMB power spectra from future experiments. In that case,
the Wishart distribution will be the correct distribution to use for 
comparing models and data on large scales.

To account for the effect of limited sky coverage on the posterior
shape, it has become common practice to model an empirically
determined posterior shape using simple fitting functions. We have
considered a number of different fits to the posterior for a single
auto-power spectrum in Section~\ref{sec:posterior_fit_single}, and
have extended these simple 1-dimensional fits to cover the combined
analysis of temperature and polarisation data. The most commonly used
fitting function for auto-power spectra is the offset log-normal
distribution. In Section~\ref{sec:posterior_fit_single}, it was shown
that this is a member of a wider class of models that form a
particular extension of the standard Gaussian posterior
form. Consequently, the extension of the log-normal distribution to
consider the combination of polarisation and temperature data is
straightforward, following the natural extension of the Gaussian
distribution to a multi-variate Gaussian. A good test of the validity
of a fitting function is its ability to match the known posterior
distribution for an all-sky survey. For the multi-variate
analogue of the log-normal distribution there is a problem, not in its
ability to fit to the posterior in the direction of a particular power
spectrum, but in fitting correlations between the different power
spectra. In fact, we have argued that this problem must be fundamental to
the class of models which adopt the same form for the posterior (in
the notation of Section~\ref{sec:posterior_fit_multi} where some
$g(C_l)$ is adopted).

An alternative approach to fitting to the power spectra is to directly
fit the pixel data. This has the advantage that the pixel values have
a multi-variate Gaussian distribution and the posterior shape is
therefore well known and simple to characterise. However, since it
requires the inversion of a $N_{\rm pix} \times N_{\rm pix}$
pixel-pixel covariance matrix, this method can only be easily employed
on CMB maps with relatively coarse pixelisation, and can therefore
only be used to probe the largest scales. To probe smaller scales,
more resolution is required and the method rapidly becomes
computationally unfeasible with increasing number of pixels. The WMAP
team in their 3-year data analysis adopted a hybrid approach where the
pixel data were directly fitted on large scales ($l\le12$ for
temperature and $l\le23$ for polarisation), and the TT and TE power
spectra were fitted on smaller scales ($l>12$ and $l>23$;
\citealt{hinshaw06,page06}). On large scales, this has the attractive
benefit of avoiding the issues arising from the complicated posterior
shape for the power spectra -- the real world complications discussed
in Section~\ref{sec:complications} do not distort the likelihood of
the pixel values from a multi-variate Gaussian distribution.

In Section~\ref{sec:marginalise} we have applied our analysis of
posterior shapes to consider marginalisation over nuisance parameters,
focusing on an unknown calibration error. Given a form for the
posterior distribution, it is possible to perform this marginalisation
analytically, leading to a simple correction to the posterior for this
``nuisance'' parameter. This analytic correction avoids having to
explicitly perform the integration. Obviously the analytic form is
dependent on the posterior shape, and it has previously been common to
assume a Gaussian posterior \citep{bridle02}. In
Section~\ref{sec:marginalise}, we provide an additional calculation
using the true likelihood of an all-sky survey. Because of the offset
nature of this distribution we find a different formula for
analytically performing the marginalisation.

The analysis presented in this paper has highlighted the issues
involved in a likelihood analysis of combined temperature and
polarisation power spectra, and provides the first step on the way to
providing a well-characterised method for the fast analysis of
combined temperature and polarisation data from future experiments. In
subsequent papers we intend to build on this work by considering the
practical application of these techniques and the possible
modifications needed for analysing upcoming joint temperature and
polarisation experiments such as the Planck experiment
\citep{tauber04}. With the precision with which future experiments
will measure the CMB temperature and polarisation fields, the likelihood 
techniques presented in this paper will become increasingly important
for accurately constraining cosmological models from these data.

\section*{Acknowledgements}
WJP and MLB are grateful for support from PPARC fellowships. We
acknowledge use of the CMBFAST \citep{seljak96} and 
HEALPIX \citep{gorski05} packages.

\setlength{\bibhang}{2.0em}

\label{lastpage}

\end{document}